\documentclass[aps,reprint,amsmath,amssymb]{revtex4-2}
\usepackage{graphicx}
\usepackage{hyperref}
\usepackage{tikz}

\begin{document}

\title{Collimated QED Cascades with Curved Plasma Mirror}

\begin{abstract}
Converting light into matter has been a longstanding goal in physics, particularly the creation of electron-positron pairs through quantum electrodynamic (QED) processes. 
While current approaches using multiple colliding laser pulses can achieve this conversion, they struggle to produce well-collimated particle beams - a crucial requirement for practical applications. 
Here we demonstrate that a single ultra-intense laser pulse, when reflected from a curved plasma mirror, can generate highly collimated electron-positron pairs with unprecedented efficiency. By focusing the laser to field strengths exceeding $a_0  > 2000$, 
our method triggers QED cascades that produce tightly focused particle beams, distinctly different from the diffuse plasmas created by conventional multi-laser setups. 
The technique works even at relatively modest laser powers of 13PW, making it immediately testable at existing facilities. 
This breakthrough opens new possibilities for studying fundamental QED processes and generating controlled matter-antimatter plasmas.
\end{abstract}

\author{X.S. Geng}
  \affiliation{State Key Laboratory of Ultra-intense laser Science and Technology,\\ 
  Shanghai Institute of Optics and Fine Mechanics (SIOM),\\
  Chinese Academy of Sciences (CAS), Shanghai, 201800, China.}

\author{M.A. Serebryakov}
\author{E.N. Nerush}
\author{A.S. Samsonov}
\author{I.Y. Kostyukov}
  \affiliation{Institute for Applied Physics of Russian Academy of Sciences, \\
  603155 Nizhni Novgorod, Russia}

\author{L.L. Ji}
  \email{jill@siom.ac.cn}
  \affiliation{State Key Laboratory of Ultra-intense laser Science and Technology,\\ 
  Shanghai Institute of Optics and Fine Mechanics (SIOM),\\
  Chinese Academy of Sciences (CAS), Shanghai, 201800, China.}

\maketitle
\section{\label{introduction}Introduction}

Pair-production, also known as electron-positron creation, is a key quantum electrodynamic (QED) process that converts light into matter~\cite{breitCollisionTwoLight1934}.
The ability to convert light into matter on demand has long been a goal of physics research.

The next-generation 100PW laser system~\cite{shaoBroadbandwidthHightemporalcontrastCarrierenvelopephasestabilized2020} offers a promising avenue for achieving this goal, 
as they are capable of generating intense electromagnetic fields that can approach or even exceed 
the Sauter-Schwinger field strength~\cite{schwingerGaugeInvarianceVacuum1951} in the rest frame of high-energy particles, 
enabling high-energy gamma radiation~\cite{ritusQuantumEffectsInteraction1985,bulaObservationNonlinearEffects1996} 
and non-linear Breit-Wheeler (BW) pair-production~\cite{reissAbsorptionLightLight1962}.
In such strong laser fields, strong-field QED cascades~\cite{bellPossibilityProlificPair2008,fedotovLimitationsAttainableIntensity2010,
elkinaQEDCascadesInduced2011,nerushLaserFieldAbsorption2011,ridgersDenseElectronPositronPlasmas2012,grismayerLaserAbsorptionQuantum2016,
zhuDenseGeVElectron2016,songDensePolarizedPositrons2022,samsonovProductionMagneticSelfconfinement2024} could develop and is the most prolific light-to-matter conversion process.
Moreover, the study of pair-production in strong laser fields also provides a unique opportunity to probe fundamental aspects of QED under extreme conditions~\cite{dipiazzaExtremelyHighintensityLaser2012,gonoskovChargedParticleMotion2022} like QED plasmas \cite{zhangRelativisticPlasmaPhysics2020}.
As such, the investigation of this important quantum process is an active area of research with far-reaching implications for both applied and basic science.

Plasma mirrors (PM) \cite{thauryPlasmaMirrorsUltrahighintensity2007,vincentiOpticalPropertiesRelativistic2014} are a promising tool for reflecting ultra-intense laser pulses.
We propose using curved PM to re-focus the incident 10-100PW laser to higher field strengths that cannot be reached with normal optical components.
An investigation \cite{vincentiAchievingExtremeLight2019} has been carried out to demonstrate this idea.
Experimentally, curved PMs have been demonstrated capable of focusing strong laser pulse to high intensities~\cite{nakatsutsumiFastFocusingShortpulse2010,arikawaUltrahighcontrastKilojouleclassPetawatt2016,wilsonEllipsoidalPlasmaMirror2016,nakatsutsumiSelfgeneratedSurfaceMagnetic2018,quereReflectingPetawattLasers2021}.
When the incident laser pulse is reflected by the curved PM, portion of the PM electrons can be extracted and accelerated by the reflected pulse~\cite{thevenetVacuumLaserAcceleration2016} to the PM focus, 
triggering the vacuum-electron-seeded-like QED cascade.
In contrast to previous researches generating cascades in standing wave formed by multiple laser pulses~\cite{bellPossibilityProlificPair2008,elkinaQEDCascadesInduced2011,zhuDenseGeVElectron2016,efimenkoExtremePlasmaStates2018,heAchievingPairCreation2022}, 
the proposed geometry uses one laser pulse, significantly reducing the complexity of experimental setup.
In our modeling, the generated pair plasma can be efficiently accelerated by the propagating wave to high energy~\cite{salaminElectronAccelerationTightly2002} rather than trapped by the standing wave, 
exhibiting high collimation and concentration, well extinguished from the background PM electrons.
The highly collimated pair plasma not only demonstrates the feasibility of large-scale light-to-matter conversion but also provides a unique platform for studying QED in extreme conditions.

\section{\label{methods}Methods}

The simulations are carried out via three-dimensional particle-in-cell code Smilei~\cite{derouillatSmileiCollaborativeOpensource2018}. 
To reflect the 100PW laser without significant distortion of the curved PM surface, 
the PM electron density is set to $1000n_c$, which corresponds to $\sim 1.7\times10^{24}$ cm$^{-3}$ at 800 nm wavelength as achieved through full ionization of heavy metals like gold, and is fully ionized in the simulation, 
which is reasonable in the interaction with focused 100PW lasers. 
To resolve the high density, the cell size is $0.01\lambda \times 0.02\lambda \times 0.02\lambda$ with $\lambda = 800$ nm being the laser wavelength. 
The simulation grid spans $2000\times 1280\times 1280$ cells with 2 macro-electrons and 1 macro-ion per cell. 
The simulation time step is 0.95 of the Courant limit, corresponding to 0.02fs. 
The surface of curved PM is a parabola of focal length of $f = 5\mathrm{\mu m}$. 
We should note that using parabola surface is convenient for modeling, 
and the interaction is not sensitive to slight surface distortion as long as the PM is able to focus. 
The incident 100PW circularly polarized (CP) laser is focused to \(w_{0} = 5\mathrm{\mu m}\) at the surface of the PM ($x = 5\mathrm{\mu m}$ in our simulation) with normalized peak field strength of \(a_{0} \approx 345\) and pulse length of 30fs. 
Here \(a_{0} = \frac{eE}{m_{e}c\omega}\), where \(e\) is the elementary charge, \(E\) the laser electric field amplitude, \(m_e\) the electron rest mass, \(c\) the speed of light in vacuum, and \(\omega = 2\pi c/\lambda\) the laser angular frequency at the fundamental wavelength \(\lambda\). 
Larger spot size is also possible and could be even better. 
We choose \(w_{0} = 5~\mathrm{\mu m}\) to save computation cost, 
since larger spot size requires larger simulation window and longer focal length.

\section{\label{results}Results}

\begin{figure}[]
  \centering
  \includegraphics[width=\columnwidth]{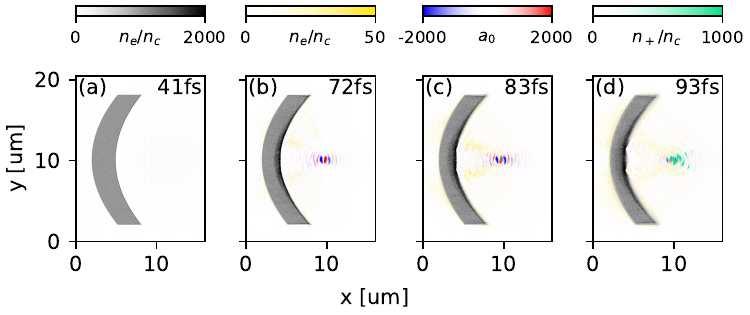}
  \caption{Evolution of the re-focusing and the generation of positrons. The electrons from the mirror with low densities are shown in yellow to show the extracted electrons. The dashed-black box indicates the $1\lambda^3$ volume near the focus.}
  \label{density-evol}
\end{figure}

Our simulations reveal a remarkable cascade process initiated when the ultra-intense laser pulse interacts with the curved plasma mirror. 
As illustrated in Fig.~\ref{ek-angled}, the process begins when the incident laser pulse reflects off the mirror (red arrows) and focuses to $x = 10~\mathrm{\mu m}$, 
achieving an extraordinary field strength of \(a_{0} > 2000\) (Fig.~\ref{density-evol}a-b). 
Simultaneously, the reflected laser extracts electrons from the mirror surface (gray trajectories in Fig.~\ref{ek-angled}), 
accelerating them toward the focus (Fig.~\ref{density-evol}b-c, shown as yellow bunches). 

Upon reaching the focus, these energetic electrons trigger the QED cascade, generating of a high-density positron beam (blue and green particles in Fig.~\ref{ek-angled} and Fig.~\ref{density-evol}d), 
with a total yield reaching 60nC --- corresponding to a 2\% conversion efficiency of the incident laser energy. 
This configuration uniquely enables a seeded cascade with signle laser pulse, 
remaining unaffected by the target, representing a novel approach to QED cascade generation.

\definecolor{ele}{rgb}{0,0.5,1}
\definecolor{pos}{rgb}{0,0.882812,0.527344}

\begin{figure}[]
  \centering
  \begin{tikzpicture}
    \node[anchor=south west] (image) at (0,0) {
      \includegraphics[width=\columnwidth]{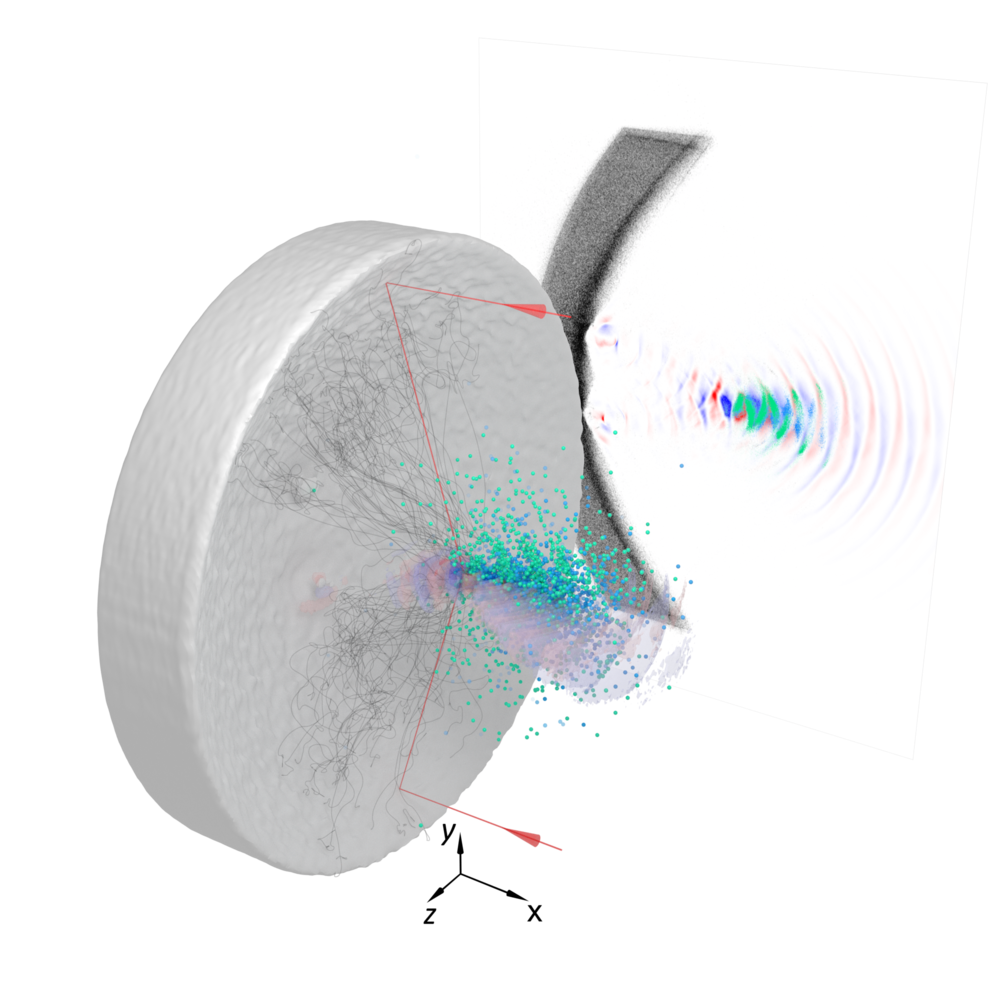}
    };
    \begin{scope}[x={(image.south east)},y={(image.north west)}]
      \draw[->, thick, gray] (0.16,0.63) -- (0.32,0.55);
      \node[gray, font=\small, anchor=south east, align=right] at (0.16,0.63) {Extracted \\[-1pt]electrons};
      
      \node[font=\small, anchor=west, fill=gray!30] at (900/2000,1100/2000) {\textcolor{ele}{e$^-$} and \textcolor{pos}{e$^+$}};
    
    \end{scope}
  \end{tikzpicture}
  \caption{Schematics of the proposed geometry. The laser pulse is reflected and re-focused to the PM focus (red arrows). The electrons are extracted from the PM and accelerated to the focus (gray lines), triggering the QED cascade (blue and green).}
  \label{ek-angled}
\end{figure}
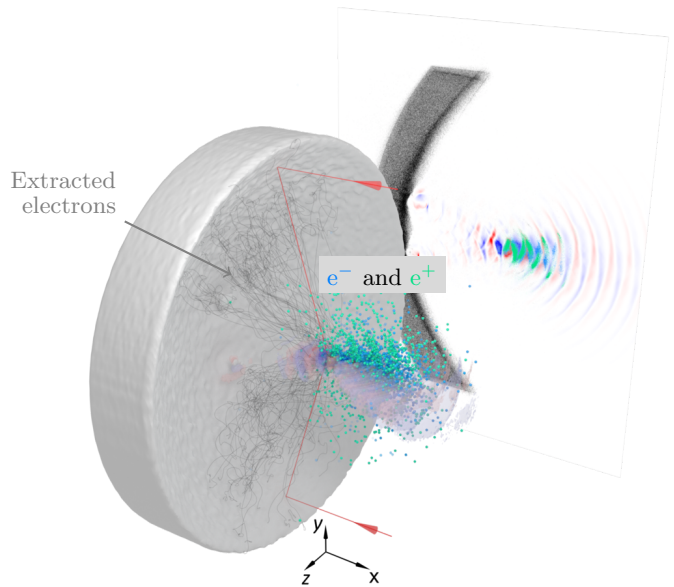

\subsection{\label{collimation-mechanism}Collimation Mechanism}

\begin{figure}[]
\centering
\includegraphics[width=\columnwidth]{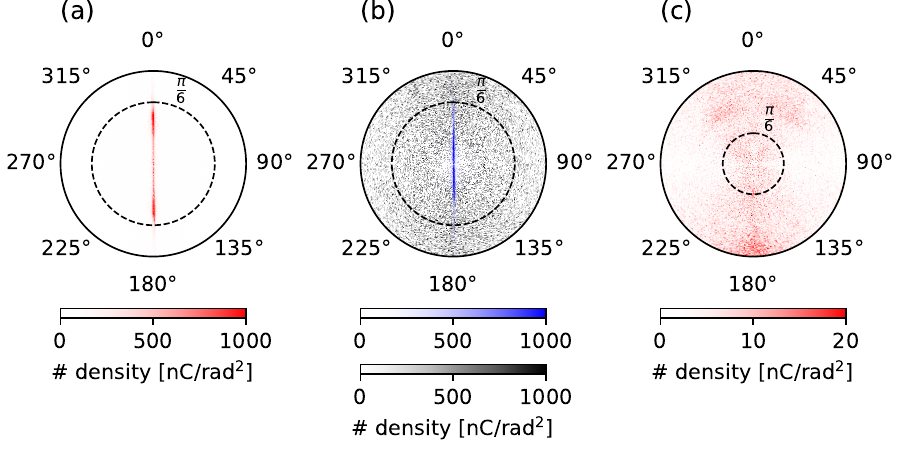}
\caption{Angular distribution of (a) the positrons and (b) the electrons (red) and background electrons (black) at 93fs. (c) The positron distribution of the flat-mirror case.}
\label{angular-dist}
\end{figure}

\begin{figure}[]
  \centering
  \includegraphics[width=\columnwidth]{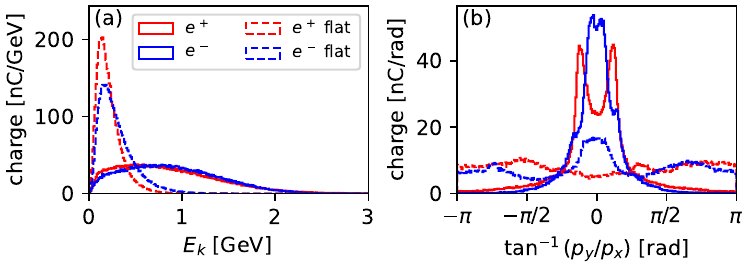}
  \caption{(a) The energy spectrums and (b) the angular distributions of the generated electrons (blue) and positrons (red) using curved PM (solid) and flat PM (dashed) at 93fs.}
  \label{ek-angle}
\end{figure}

Beyond the high positron yield, our simulations reveal another key advantage: the generated pair-plasma exhibits exceptional collimation (Fig.~\ref{angular-dist}a-b), creating high contrast against the background PM electrons. 
This collimation stems from the unique field configuration ---
while the incident laser field is insufficient for triggering QED cascades, the refocused pulse reaches \(a_{0} > 2000\), much higher than the incident laser, 
enabling the plasma to form within a propagating wave rather than a standing wave~\cite{bellPossibilityProlificPair2008,elkinaQEDCascadesInduced2011,zhuDenseGeVElectron2016,heAchievingPairCreation2022}. 
The intense fields efficiently accelerate both electrons and positrons to high energies~\cite{salaminElectronAccelerationTightly2002}, producing a concentrated angular distribution characteristic of linear polarization (Fig.~\ref{ek-angle}a).

To highlight the significance of our curved PM geometry, we performed comparative simulations using a flat PM with identical laser power but tightly focused to $w_0=2\mathrm{\mu m}$ onto the flat PM. 
As shown in Fig.~\ref{angular-dist}c, the flat PM configuration generates pair plasma within counter-propagating waves, severely limiting particle acceleration. The energy spectrums are compared in Fig.~\ref{ek-angle}a. 
The resulting angular distribution in Fig.~\ref{angular-dist}c shows only minimal polarization-direction anisotropy, dramatically contrasting with our curved PM results.

Another noteworthy feature in Fig.~\ref{angular-dist}(a) is the dual-spike of the generated positrons, which is also presented in Fig.~\ref{ek-angle}(b). 
In general, electrons and positrons should exhibit symmetric distributions in the laser fields. 
The observed asymmetric positron distribution arises from their distinct scattering dynamics.
The dual-spike feature happens in both the curved and flat PM setups. 

\begin{figure}[h]
  \centering
  \includegraphics[width=\columnwidth]{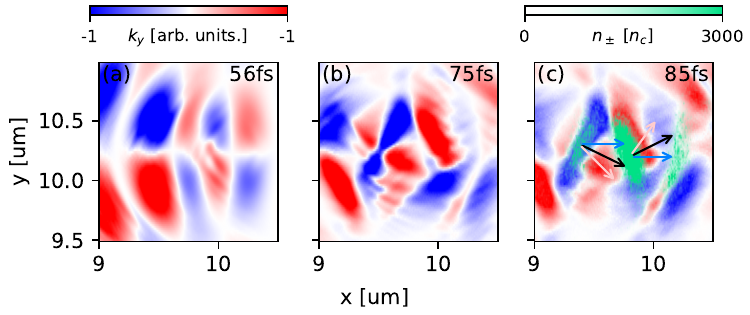}
  \caption{
    Evolution of the distribution of the laser wave vector \(k_{y}\) near the focus at (a) 56fs, (b) 75fs and (c) 85fs. 
    In (c) the black arrow denotes the light pressure direction of the laser fields. 
    The pink and blue arrows denote the scattering directions of positrons and electrons of the pair plasma.}
  \label{pair-in-ky}
\end{figure}

According to our modelling, the effect arises from the asymmetry of the re-focused laser due to distortion of the PM surface.
To highlight the effect, the wave vector \(k_{y} = E_{z}B_{x} - E_{x}B_{z}\) is shown in Fig.~\ref{pair-in-ky}(a-c). 
At 56fs, before the re-focused \(k_{y}\) is symmetric (Fig.~\ref{pair-in-ky}(a)) since this part is the rising edge of the laser pulse and the PM surface is not yet distorted.
At 75fs, the \( k_{y}\) becomes rather distorted (Fig.~\ref{pair-in-ky}(b)) since the PM surface is distorted by the laser peak (\(a_0 \approx 345\)).

Later at 85fs, the pair plasma is generated and is shown in Fig.~\ref{pair-in-ky}(c), where the balck arrows denote the local wave vector of the laser.
In the asymmetric \(k_y\) field, the generated pair plasmas in adjacent laser phases experience opposite light pressure as denoted by the blue arrows. 
The light pressure direction is mostly along the \emph{x} direction, but a small \(k_{y}\) emerges in the tightly focused area as shown in Fig.~\ref{pair-in-ky}(c). 
Then, electrons and positrons in the same laser phase will get scattered to opposite directions due to opposite charge, denoted by the blue and pink arrows. 
Finally, the electrons are mostly scattered along the \emph{x} direction, and the positrons are scattered off the \emph{x} direction, forming the single-spike and dual-spike structures in Fig.~\ref{ek-angle}(b).

\begin{figure}[h]
  \centering
  \includegraphics[width=\columnwidth]{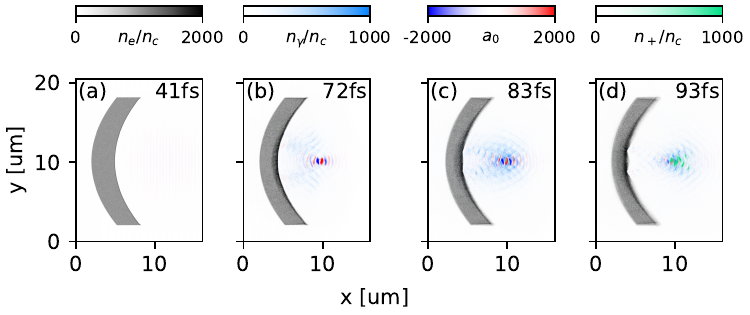}
  \caption{Same as Fig.~\ref{density-evol} with the density of gamma photons shown by blue areas.}
  \label{gamma-density-evol}
\end{figure}

On the other hand, significant amount of gamma photons is also generated and focused to the laser focus, as shown in Fig.~\ref{gamma-density-evol}. 
Here comes the question whether the pair-production is driven by photons or by electrons since the maximum density is more than 10 times of the extracted electrons. 

\subsection{\label{photon-pair-analysis}Photon-Pair Conversion Analysis}

\begin{figure}[h]
  \centering
  \includegraphics[width=\columnwidth]{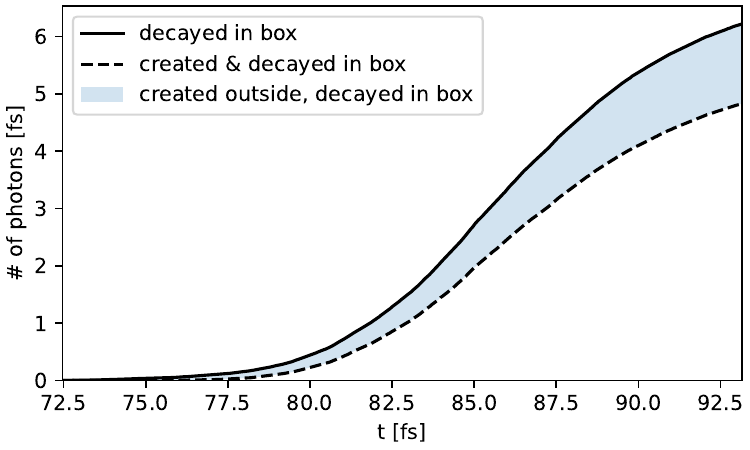}
  \caption{Evolution of the number of photons decayed in the box (black-solid), photons created and decayed in the box (black-dashed) and their difference, i.e., photons created outside but decayed in the box (blue area). The region of the box is defined in Fig.~\ref{gamma-density-evol}(d).}
  \label{photon-decay}
\end{figure}

Therefore, we need to infer how many pairs are produced by the incident gamma photons and how many are created by the extracted electrons through cascading. 
Direct calculation of this problem is impossible since the pairs produced by the gamma photons and the extracted electrons are indistinguishable, where the former are also radiated by the latter before entering the focusing region. 
We propose evaluating the number of the photons decayed near the focus and the photons created and decayed near the focus, as shown in Fig.~\ref{photon-decay}, where the region near focus is defined by a box in Fig.~\ref{gamma-density-evol}(d). 
The difference between the two is therefore the number of photons created far from the focus but decayed near the focus, which can be seen as the number of pairs driven by the gamma photons from the mirror, which makes up 1/4 of the decayed photons, i.e., 1/4 of the first-generation pairs. 
Therefore, the QED cascade in triggered and driven by mainly the extracted electrons from the PM.

\begin{figure}[]
  \centering
  \includegraphics[width=\columnwidth]{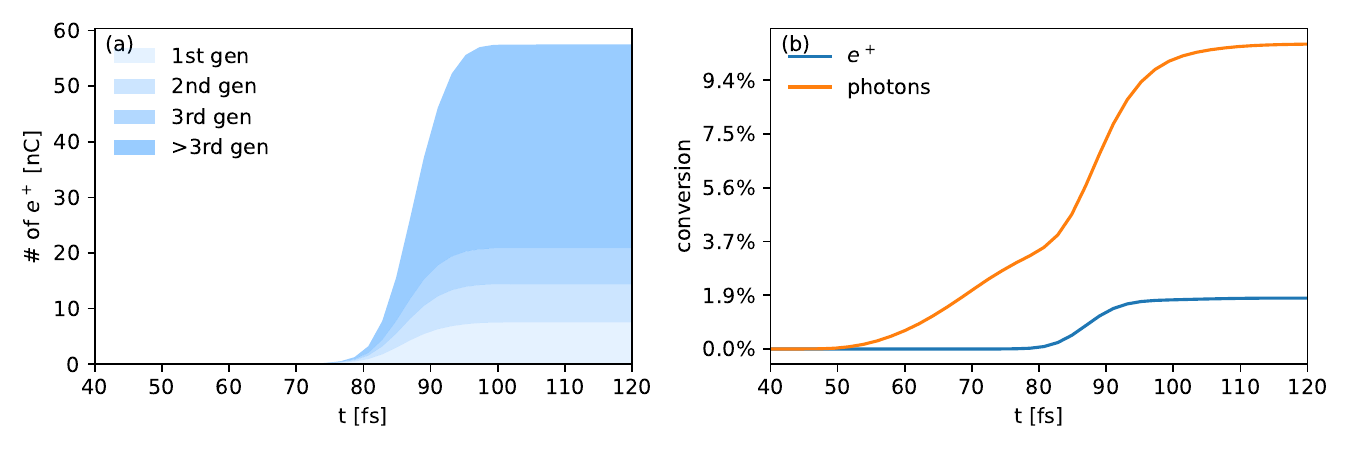}
  \caption{(a) Evolution of the number of the positrons for 1st to 3rd and greater than 3rd generations. (b) Evolution of the conversion efficiency from laser to photons and positrons.}
  \label{efficiency}
\end{figure}

We estimate the generations of the produced pairs for demonstration of the avalanche-type cascade \cite{bellPossibilityProlificPair2008}, i.e., the photon \(\rightarrow\) pair \(\rightarrow\) photon loop should repeat many times.
In our modelling, more than 3 generations are observed \footnote{We use different species \emph{ele\_gen1}, \emph{ele\_gen2}, \emph{ele\_gen3} and \emph{ele\_gt3} to denote the generations, same to the positrons and photons. Too many kinds of species will induce performance issue. Therefore we only show the first 3 generations.} in Fig.~\ref{efficiency}(a) and at least 7 generations can be estimated according to the per-generation number in Fig.~\ref{efficiency}(a), indicating the avalanche type of QED cascade.

\begin{figure}[h]
  \centering
  \includegraphics[width=\columnwidth]{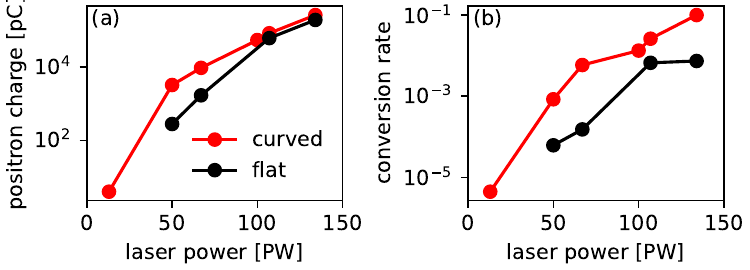}
  \caption{
    (a) The positron yields and 
    (b) the conversion efficiency from laser to positrons for the curved PMs (red-dotted) and flat PMs (black-dotted), 
    with laser power ranging from 13PW to 137PW.}
  \label{power-scale}
\end{figure}

Next, we explore the positron yields at higher and lower laser powers other than the 100PW case. 
As the laser power increases, the positron yields grow exponentially with laser power above 50PW in both the curved and flat PM cases as shown in Fig.~\ref{power-scale}(a), and the total yields are close at high powers. 
In the flat PM case, the laser is tightly focused to $2\mu \text{m}$ on the PM surface. 
It should be noted that although the positron yields are close at high powers, the collimation of the pair can be significantly different, as discussed previously.
The curved PM geometry can even generate a few pC of positrons at 13PW, which can be tested on 10PW-class laser systems~\cite{dansonPetawattExawattClass2019a}, 
although these positrons are not generated through QED cascading but solely BW process due to insufficient laser intensity.
In terms of conversion efficiency, the efficiency of the flat-PM geometry is much lower than the curved-PM geometry, which is due to the insufficient acceleration in the colliding pulses as discussed before. 
The positrons can even absorb 10\% of the incident laser power at 137PW, which is within expectation since pair plasma generated in standing wave can even drain the laser energy at high powers~\cite{fedotovLimitationsAttainableIntensity2010,nerushLaserFieldAbsorption2011,wuUpperLimitLaser2021}.

\section{\label{discussion}Discussion}

\begin{figure}[h]
  \centering
  \includegraphics[width=\columnwidth]{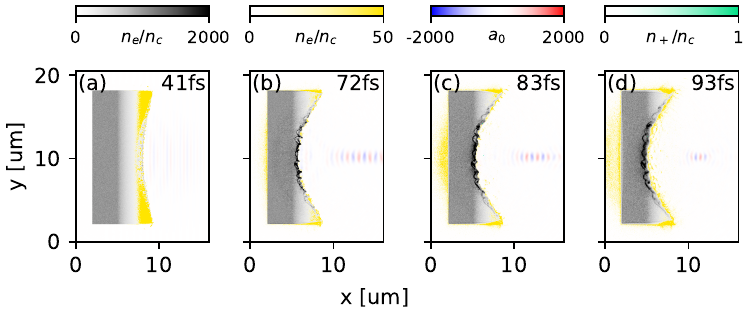}
  \caption{Interaction of 100PW laser with flat PM with preplasma. A curved surface is formed by light pressure and the laser is well re-focused, but is unable to produce pairs.}
  \label{preplasma}
\end{figure}

In this section we discuss the possibility of creating a curved PM through the denting of preplasma induced by the incident laser pulse~\cite{vincentiAchievingExtremeLight2019,vincentiOpticalPropertiesRelativistic2014}, 
where the denting of preplasma is dependent on the laser field strength and a curved surface can be formed due to the Gaussian spot. 
In Fig.~\ref{preplasma}, we investigate the process with a 100PW laser and a flat PM with preplasma of $l_\mathrm{pre}=1.5~\mathrm{\mu m}$. 
As shown in Fig.~\ref{preplasma}, the curved PM is well formed and the reflected laser is tightly focused to high field like the results in Fig.~\ref{density-evol}. 
However, no pairs are generated near the focus, since no electrons are extracted to the focus (blue area). 
Electrons in the preplasma experience large acceleration gradient from laser due to the low plasma density and almost all electrons in the preplasma are accelerated towards the laser incident direction. 
Therefore, another weaker pulse before the main pulse is necessary to dent the preplasma without significant acceleration, 
since the reflection of electrons are the key for triggering the QED cascades as shown by our results.

\section{\label{conclusion}Conclusion}

This work demonstrates a significant breakthrough in light-to-matter conversion using curved plasma mirrors. 
Our approach solves a fundamental challenge in QED cascade generation - achieving highly collimated electron-positron beams from laser-plasma interactions. 
By focusing 100PW-class lasers to unprecedented field strengths ($a_0 > 2000$), we generate tightly focused pair plasmas with efficiency and beam quality far superior to conventional multi-laser methods. 
The remarkable collimation and high contrast of the generated positron beams represent a crucial step toward practical applications in high-energy physics. 
Most significantly, our demonstration of positron generation at just 13PW brings these exotic QED processes within reach of existing laser facilities, opening immediate opportunities for experimental validation. 
This accessible path to studying extreme QED phenomena could accelerate our understanding of fundamental physics and advance technologies for future matter-antimatter colliders and novel radiation sources.

\acknowledgments{
  This work is supported by the National Key R\&D Program of China (No. 2022YFE0204800), 
  the National Natural Science Foundation of China (Nos. 12388102, 12374298 and 12304384) and
  the Strategic Priority Research Program of Chinese Academy of Sciences (No. XDB890303).
}
\bibliography{refs}
\end{document}